\documentstyle{mn}
\input{epsf}

\title[Spherical infall model]{Universal profile of dark matter halos and
the spherical infall model}
\author[Ewa L. {\L}okas]{Ewa L. {\L}okas\\ Copernicus Astronomical Center,
Bartycka 18, 00--716 Warsaw, Poland}

\begin{document}

\maketitle

\begin{abstract}
I propose a modification of the spherical infall model for the evolution
of density fluctuations with initially Gaussian probability distribution
and scale-free power spectra in Einstein-de~Sitter universe as developed by
Hoffman \& Shaham. I introduce a generalized form of the initial density
distribution around an overdense region and cut it off at half the
inter-peak separation accounting in this way for the presence of the
neighbouring fluctuations. Contrary to the original predictions of Hoffman
\& Shaham the resulting density profiles within virial radii no longer have
power-law shape but their steepness increases with distance. The profiles
of halos of galactic mass are well fitted by the universal profile formula
of changing slope obtained as a result of $N$-body simulations by Navarro,
Frenk \& White. The trend of steeper profiles for smaller masses and
higher spectral indices is also reproduced. The agreement between the
model and simulations is better for smaller masses and lower spectral
indices which suggests that galaxies form mainly by accretion while
formation of clusters involves merging.

\end{abstract}


\section{Introduction}

It is generally believed that the structure in the universe emerged
hierarchically from initially small density fluctuations. Small
fluctuations, which at present remain such only at very large smoothing
scales, can be successfully described by the linear theory. On smaller
scales however, nonlinear effects come into play and linear approximation
is no longer valid. In the weakly nonlinear regime perturbative approach
proved extremely successful in predicting the statistical properties of
density fields at scales exceeding 10 $h^{-1}$ Mpc. In order to predict
properties of single objects we must however trace the evolution of
density up to strongly nonlinear regime. The price to pay is high: we have
to restrict the analysis to one object neglecting its interactions with
the neighbourhood.

The simplest deterministic model of strongly nonlinear evolution proposed
by Gunn \& Gott (1972), called the spherical model, described the
behaviour of a uniformly overdense region in the otherwise unperturbed,
expanding Universe. It was extended by Gott (1975) and Gunn (1977) to
apply to the evolution of matter around an already collapsed density
perturbation superposed on a homogeneous background. The main prediction
of the model (called the spherical accretion or the secondary infall
model, hereafter SIM) was that the matter collapsing around the
perturbation should form a halo with $r^{-9/4}$ density profile.

It is much more realistic to assume that the progenitors of structure were
not the collapsed perturbations but the local maxima (rare events) in the
density field which had initially Gaussian probability distribution. This
was the approach of Hoffman \& Shaham (1985, hereafter HS) who applied
SIM to the hierarchical clustering scenario. They assumed that the
density peak dominates to some extent the surroundings causing the
collapse of the material that is gravitationally bound to it. The initial
density profile around the peak was approximated by the mean density
predicted by the initial probability distribution which was assumed to be
Gaussian. HS considered scale-free initial power spectra of fluctuations
in different cosmologies and found that the final profiles of halos
steepen for higher spectral indices and lower density parameter
$\Omega_{0}$.

This kind of dependence on cosmological parameters was confirmed via the
results of completely different approach to studies of structure
formation: the $N$-body simulations. Quinn, Salmon \& Zurek (1986),
Efstathiou et al. (1988) and more recently Crone, Evrard \& Richstone
(1994) among others confirmed the analytical predictions by fitting power
laws to the density profiles of dark matter halos resulting from their
simulations.

Recently Navarro, Frenk \& White (1997, hereafter NFW) performed a series
of high-resolution $N$-body simulations for power-law initial spectra and
found that the density profiles of dark halos in large range of masses can
be fitted with a simple formula with only one fitting parameter. The
density profile was observed to steepen from $r^{-1}$ near the
centre of the halo to $r^{-3}$ at large distances. This confirmed earlier
results of these authors obtained for CDM cosmologies. Similar shapes of
profiles were also observed by Cole \& Lacey (1996) and Tormen, Bouchet \&
White (1997) in their $N$-body simulations.

Although the overall trend of steeper profiles for higher spectral indices
and lower $\Omega_{0}$ predicted by SIM was confirmed, NFW claim that the
changing slope of the profile is ``at odds with the analytic predictions".
In fact this is only true in the case of $\Omega_{0}=1$ for which SIM
indeed predicts a power-law profile. However, for large (i.e. close to 1)
values of $\Omega_{0}$ SIM in the form proposed by HS is least reliable.
This is mainly due to the fact that SIM describes the evolution of a
single overdense region while in reality the halos forming at various
locations compete for mass. As a result, the mass accreted by an overdense
region is not the whole mass which is gravitationally bound to it (in the
case of $\Omega_{0}=1$ this mass is infinite). Another limitation of
SIM comes from the statistical approach applied in determining the
initial conditions: the initial density distribution has a finite
coherence scale which also may influence the amount of mass accreted onto
the peak.

The paper is organized as follows: after a short presentation of
SIM in Section~2, in Section~3 and 4 I outline proposed modifications to
the model and in Section~5 compare obtained halo density profiles to
the results of $N$-body simulations. The discussion follows in Section~6.

\section{The standard spherical infall model for $\Omega=1$}

In the following I present the main assumptions and results of
SIM as applied to the density contrast field with
initially Gaussian probability distribution. This version of the model was
first presented for the case of linear fields by HS
and slightly modified by {\L}okas (1998) in attempt to take into account
the weakly nonlinear corrections. In the following only the linear
approximation will be used to determine initial conditions. The remaining
assumptions and conventions will follow those of \L okas (1998)
except for a change in notation introduced in order to conform to the
widely accepted notation for the parameters of the universal profile.

I will consider the Einstein-de~Sitter cosmological model with $\Omega=1$
and zero cosmological constant. The initial probability distribution of
the density field will be assumed to be Gaussian. The initial density
power spectrum will be modeled by the scale-free form
\begin{equation}  \label{p1}
    P(k) = C k^n
\end{equation}
with indices $n=-1.5, -1, -0.5, 0$ which are of biggest cosmological
interest. The fields will be smoothed with a filter of a Gaussian
shape
\begin{equation}  \label{p2}
    W_{\rm G}(kR)={\rm e}^{-k^{2} R^{2}/2}.
\end{equation}
For such spectra and smoothing the density field can be characterized by
the correlation coefficient
\begin{equation}    \label{p3}
    \varrho = \frac{\xi_{R}(r)}{\sigma^2} = \ _{1}F_{1} \left(
    \frac{n+3}{2}, \frac{3}{2}, - \frac{x^2}{4} \right)
\end{equation}
where $\xi_{R}(r)$ is the (auto)correlation function of the two values of
the density fields (smoothed at comoving scale $R$) at points separated by
distance $r$, $x$ is the distance in units of $R$: $x=r/R$ and $\sigma$ is
the linear rms fluctuation at scale $R$ which in this case is given by
\begin{equation}   \label{p4}
    \sigma^{2} = C D^{2}(t) \frac{\Gamma[(n+3)/2]}{(2 \pi)^{2} R^{n+3}}
\end{equation}
where the time dependence is
\begin{equation}    \label{p4a}
    D(t) = \frac{1}{1+z} .
\end{equation}

Let us assume that at $r=0$ we detect an overdense region of density equal
to $a$ standard deviations: $\delta(r=0) = a \sigma$ (for high enough
$a$ this is approximately equivalent to assuming there is a peak at $r=0$).
Two-point probability distribution function then predicts that at distance
$r$ (or $x$ if we measure the distance in units of $R$) the expected
density contrast is
\begin{equation}        \label{p5}
        \langle \delta(x) \rangle = \varrho(x) a \sigma
\end{equation}
with $\varrho(x)$ given by equation (\ref{p3}).

The dynamical evolution of matter at the distance $x_{\rm i}$ from the peak
is determined by the mean cumulative density perturbation within
$x_{\rm i}$ which is given by
\begin{equation}  \label{p6}
    \Delta_{\rm i}(x_{\rm i}) = \frac{3}{x_{\rm i}^{3}}
    \int_{0}^{x_{\rm i}} \delta(x) x^{2} {\rm d} x.
\end{equation}
If we assume that $\delta(x) = \langle \delta(x) \rangle$ the expected
cumulative density can be approximated by
\begin{equation}           \label{p8}
    \langle \Delta_{\rm i}(x_{\rm i}) \rangle = \left\{
\begin{array}{ll}
       a \sigma &\ \ {\rm for} \ \ x_{\rm i} \ll 1 \\
       a \sigma h(n)x_{\rm i}^{-(n+3)}   &\ \ {\rm for}
    \ \ x_{\rm i} \gg 1.
\end{array} \right.
\end{equation}
The values of the numerical factor $h(n)$ can be found in \L okas (1998).
A similar approximation for large $x_{\rm i}$ was used by HS as the
initial condition for their SIM. The approximation
at large $x_{\rm i}$ in the form given above is accurate to within 10\%
for $x_{\rm i} \ge 5$.

The cumulative density contrast $\Delta_{\rm i}(x_{\rm i})$ describes the
initial density distribution around the peak. Assuming that the
cumulative density can be approximated by the mean, $\Delta_{\rm i}(x_{\rm
i})= \langle \Delta_{\rm i}(x_{\rm i}) \rangle $, SIM can be used to
predict the final density profile which will form after shells which are
bound to the peak collapse onto it. The distance to the farthest shell
which is bound to collapse is given by the condition of the vanishing
energy \begin{equation}    \label{p9} \Delta_{\rm i}(x_{\rm i,0}) =
    \Omega_{\rm i}^{-1} - 1 = 0 \end{equation} where $\Omega_{\rm i}$ is
the density parameter at some initial epoch at which we specify the
initial density distribution. In the case of $\Omega=1$ the scale $x_{\rm
i,0}$ of the gravitational influence of any overdense region is infinite,
the region should therefore collect mass from the whole universe.

According to SIM after virialization the shell
$x_{\rm i}$ ends up at $x=f x_{\rm m}$ where $x_{\rm m}$ is the maximum
radius given by
\begin{equation}    \label{p9a}
    x_{\rm m} = x_{\rm i} \frac{\Delta_{\rm i} + 1}{\Delta_{\rm i}}
\end{equation}
and $f$ is the collapse factor, which in the case of scale free initial
conditions for large $x_{\rm i}$, equation (\ref{p8}), can be approximated
by a constant of the order of $1/2$ (see Zaroubi \& Hoffman 1993). The
final density profile then follows from the conservation of mass
\begin{equation}  \label{p9b}
    \rho(x) x^{2} {\rm d} x = \rho_{\rm i}(x_{\rm i}) x_{\rm i}^{2}
    {\rm d} x_{\rm i}.
\end{equation}
If we approximate the initial density of the shell of radius $x_{\rm i}$
\begin{equation}  \label{p9c}
    \rho_{\rm i}(x_{\rm i}) = \rho_{\rm b,i} [1 + \delta_{\rm i}
    (x_{\rm i})]
\end{equation}
by the background (critical) density
$\rho_{\rm b,i}=\rho_{\rm crit,i}=(1+z_{\rm i})^{3} \rho_{\rm crit,0}$
and expand the right hand side of equation (\ref{p9a}) in $\Delta_{\rm
i}$ keeping only the linear term, we will end up with the power law density
profile found by HS
\begin{equation}    \label{p9d}
    \frac{\rho (x)}{\rho_{\rm crit,0}} = \frac{(1+z_{\rm i})^3}{n+4}
    \left( \frac{a \sigma h}{f} \right)^{3/(n+4)} x^{-3(n+3)/(n+4)}.
\end{equation}
The density profiles will be expressed here as above in units of the
present critical density to ensure direct comparison with the results of
$N$-body simulations.

It is interesting to note that what is usually quoted as the prediction of
SIM for hierarchical clustering scenarios is the
special case (\ref{p9d}) which is valid only for $\Omega=1$. In the case
of open universe the slope of the halo steepens gently with the distance
from the centre starting with the slope (\ref{p9d}) near the centre and
approaching $x^{-4}$ for shells close to the last bound shell. The case of
$\Omega<1$ will be treated in the follow-up paper.

\section{How to improve the spherical infall model?}

The main and most questionable assumptions underlying SIM as formulated by
HS are of course the spherical symmetry of the problem and the absence of
peculiar velocities. I will keep the assumptions here and show that even
with these simplifications the agreement between the model and the results
of $N$-body simulations can be improved.

First I propose to relinquish the large $x_{\rm i}$ approximation for the
$\langle \Delta_{\rm i}(x_{\rm i}) \rangle$, equation (\ref{p8}) applied
by HS in their calculations. The general formula for the expected
cumulative density contrast can be found using equations (\ref{p3}),
(\ref{p5}) and (\ref{p6})
\begin{eqnarray}
    \langle \Delta_{\rm i}(x_{\rm i})  \rangle &=& \frac{6 a \sigma}{(n+1)
    x_{\rm i}^{2}}  \left[ \
    _{1}F_{1} \left( \frac{n+1}{2}, \frac{3}{2}, - \frac{x_{\rm i}^{2}}{4}
    \right) \right. \nonumber\\ &-& \left. \ _{1}F_{1} \left(
    \frac{n+1}{2}, \frac{1}{2}, - \frac{x_{\rm i}^{2}}{4} \right) \right].
    \label{p7}
\end{eqnarray}

The main reason for this generalization is that especially with the
conditions given below the main contribution to the final density profile
of the halo comes from $x_{\rm i}$ of the order of few. Although HS used
the large $x_{\rm i}$ approximation as the one applicable at large
distances from the peak, it is possible to interpret large $x_{\rm i}$
also as small smoothing scales. The smoothing procedure of the initial
density field that is performed here in order to determine the location of
the overdense regions has no equivalent in real structure formation or
$N$-body simulations where the only artificial scales are those of
softening length and the size of the simulation box. One may argue that
the large distance (or small smoothing scale) approximation to $\langle
\Delta_{\rm i}(x_{\rm i}) \rangle$ is therefore more realistic. However,
taking into account the intrinsic role of the initial smoothing in
determination of the initial conditions for collapse, the masses of halos
etc. it is difficult to accept such argument.

\begin{figure}
\begin{center}
    \leavevmode
    \epsfxsize=8cm
    \epsfbox[96 77 372 353]{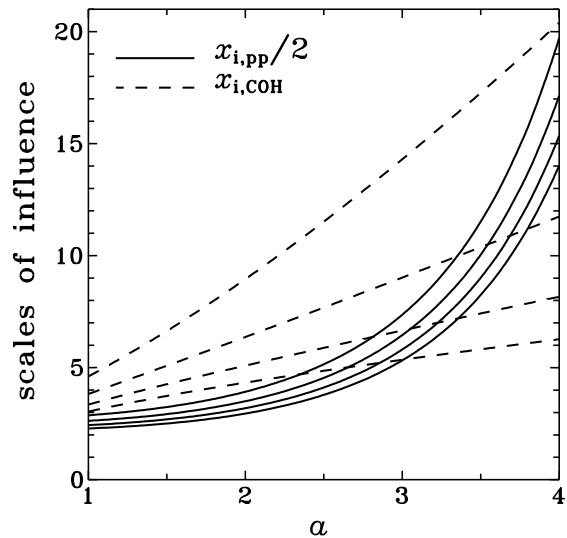}
\end{center}
    \caption{The scale of coherence $x_{\rm i,COH}$ (dashed lines) and half
    the distance between peaks $x_{\rm i,pp}/2$ (solid lines) as functions of
    the height of the peak $a$ for different spectral indices. The
    four curves in each set correspond from top to bottom to $n=-1.5, -1,
    -0.5$ and $n=0$.}
\label{ccohcpp}
\end{figure}

Another limitation of the model presented in Section~2 comes from the
statistical approach to determining the initial conditions for collapse.
The density profile is evaluated with the assumption $\Delta_{\rm
i}(x_{\rm i}) = \langle \Delta_{\rm i}(x_{\rm i}) \rangle$, while this
approximation can only be considered valid for scales up to the scale of
coherence $x_{\rm i,COH}$ defined by
\begin{equation}    \label{p31}
    \langle \Delta_{\rm i} \rangle  = \Sigma_{\Delta}.
\end{equation}
The calculation of the dispersion $\Sigma_{\Delta}$ in the case of
power-law spectra and Gaussian smoothing is presented in Appendix~A.
Equation (\ref{p31}) can be then solved numerically for $x_{\rm i,COH}$.
The results for different spectral indices $n$ and heights of the peak $a$
are shown in Figure~\ref{ccohcpp}. Table~\ref{tabc} lists
the values of $x_{\rm i,COH}$ for the peak of height $a=3$ that will be
considered in the calculation of the profiles.

\begin{table}
\caption{The values of the scales of influence $x_{\rm i,COH}$ and $x_{\rm
  i,pp}/2$ for peaks of height $a=3$ for different spectral indices $n$.}
\label{tabc}
\begin{tabular}{rlc} $n$ & $x_{\rm i,COH}$ & $x_{\rm i,pp}/2$\\
  \hline
-1.5 & 14.3 & 7.36 \\
-1.0 & $\ \, 9.02$ & 6.45 \\
-0.5 & $\ \, 6.66$ & 5.81 \\
 0.0 & $\ \, 5.35$ & 5.32
\end{tabular}
\end{table}

Similar calculations, but without providing the analytical expressions for
$\langle \Delta_{\rm i} \rangle$ and $\Sigma_{\Delta}$, were performed by
HS and also by Peebles (1984) and Ryden (1988) who considered peaks
instead of overdense regions. HS and Ryden (1988) then consider the
fraction of mass subscribed to peaks higher than $a \sigma$ in the case
of different power spectra
\begin{equation}    \label{p32}
    F( >a) = \frac{4 \pi}{3} \int_{a}^{\infty} [r_{\rm i,COH}(\nu)]^{3}
    N_{\rm pk}(\nu) {\rm d} \nu
\end{equation}
where $N_{\rm pk}(\nu)$ is the differential number density of peaks given
by equations (4.3)-(4.10b) of Bardeen et al. (1986, hereafter BBKS). Since
$r_{\rm i,COH}(\nu)$ grows faster with $\nu$ for lower spectral
indices, for example $F(>3)$ is below unity for higher spectral indices to
a few for lower spectral indices. This leads HS to assume that the most
probable peaks around which structure will form are those for which
$F(>a)=1$. However, this way they end up with a rather surprising value
of $a>4$ for the height of the most probable peak for $n=-2$. From the
statistics of peaks by BBKS we know that $N_{\rm pk}(\nu)$ is a strongly
decreasing function for large $\nu$ and peaks with $a>4$ are very rare.

I propose here to consider the distance between peaks as an additional
constraint on the cumulative density distribution around a peak. It is
reasonable to assume that only the peaks higher than $a$ should be
considered as ``dangerous" for the structure forming around our chosen
peak. Since the number density of peaks higher than $a$ for
sufficiently high values of $a$ ($a=2,3,4$) is a strongly decreasing
function of $a$, we may assume that the peaks nearest to our peak are of
height close to $a$. The scales of influence of our peak and the
neighbouring one will be similar and we may approximate this scale as half
the distance between peaks, $x_{\rm i,pp}/2$. The scale $x_{\rm i,pp}$ is
determined by the number density of peaks higher than $a$:
\begin{equation}     \label{p33}
    x_{\rm i,pp} = \frac{[n_{\rm pk} (>a)]^{-1/3}}{R}
\end{equation}
where
\begin{equation}    \label{p34}
    n_{\rm pk} (>a) = \int_{a}^{\infty} N_{\rm pk}(\nu) {\rm d} \nu.
\end{equation}

The numerical results for $x_{\rm i,pp}/2$ for different spectral indices
are shown in Figure~\ref{ccohcpp}. The results for $a=3$ are also listed
in Table~\ref{tabc}. In the following I will simulate the constraints on
the cumulative density (\ref{p7}) by cutting it off at the smaller of the
two scales $x_{\rm i,pp}/2$ and $x_{\rm i,COH}$. I have chosen to consider
here peaks of given height $a=3$ which are high enough to collapse by
themselves and frequent enough to produce large number of objects. We
see that the scale $x_{\rm i,pp}/2$ for $a=3$ is always smaller than the
corresponding $x_{\rm i,COH}$ except for the highest spectral index
considered, $n=0$, where the two scales are almost equal. This motivates
the introduction of the cut-off at $x_{\rm i,pp}/2$ and not $x_{\rm
i,COH}$.

The remaining issue is the shape of the cut-off function. I will adopt a
sharp cut-off such that the corrected cumulative density will be
\begin{equation}    \label{p35}
    \Delta_{\rm i,cut}(x_{\rm i}) = \frac{\langle \Delta_{\rm i}(x_{\rm i})
    \rangle} {1+{\rm e}^{(x_{\rm i} - x_{\rm i,pp}/2)/w}}
\end{equation}
with $\langle \Delta_{\rm i}(x_{\rm i}) \rangle$ given by equation
(\ref{p7}). Unfortunately, in contrast to the problem of the scale of
cut-off, we cannot predict the width of the filter, $w$ from the model. It
will be treated as an unknown parameter which has to be specified by
matching the results from SIM and $N$-body simulations (see Section~5).

Summarizing, the final density profile of the virialized halo is given by
\begin{equation}    \label{p36}
    \frac{\rho (x)}{\rho_{\rm crit,0}} = (1 + a \sigma \varrho) (1+ z_{\rm
    i})^3 \left(\frac{x_{\rm i}}{x} \right)^2 \frac{{\rm d} x_{\rm
    i}}{{\rm d} x}
\end{equation}
with
\begin{equation}    \label{p37}
    x = \frac{x_{\rm i} f [\Delta_{\rm i,cut}(x_{\rm i}) +1]}
    {\Delta_{\rm i,cut}(x_{\rm i})}.
\end{equation}
Formula (\ref{p36}) gives a complicated but analytical expression for the
density profile as a function of the initial radius $x_{\rm i}$ which is
related to the final radius $x$ by equation (\ref{p37}). Since the initial
density distribution in its generalized form (\ref{p7}) is not scale free
the collapse factor $f$ in equation (\ref{p37}) can no longer be assumed
to be the same constant for all shells (for the results with $f=1/2$ see
\L okas 1999). Detailed calculation of this factor is presented in the
following section.

\section{The collapse factor}

The purpose of this section if to find a correction to the fiducial
density profile defined as the density distribution with all shells
stopping at their maximal radii. This corresponds to putting $f=1$ in
equations (\ref{p36})-(\ref{p37}). In reality after reaching the maximum
radius a given shell will start oscillating and it will contribute to the
actual mass contained inside inner shells which are expected
to contract adiabatically (see the discussion in Zaroubi \& Hoffman
1993) by a factor
\begin{equation}   \label{f1}
    f(x_{\rm i}) = \frac{M(x_{\rm i})}{M(x_{\rm i}) + m_{\rm add}(x_{\rm
    i})}
\end{equation}
where $M(x_{\rm i})=M(x)$ is the mass contained inside the shell of radius
$x_{\rm i}$ or $x$ and $m_{\rm add}(x_{\rm i})$ is the mass contributed
by the outer shells.

If the shape of the fiducial profile is a power-law given by (\ref{p9d})
the collapse factor is easily determined to be
\begin{equation}    \label{f2}
    f = \frac{1}{1+ (3-\gamma) \int_{1}^{U} u^{2-\gamma} P(1/u) {\rm d} u}
\end{equation}
where $\gamma = 3(n+3)/(n+4)$, $U$ is the size of the system in units of
the radius of the considered shell, $P$ is the fraction of time a particle
belonging to shell of radius $x'$ spends within radius $x$
\begin{equation}    \label{f3}
    P(x, x') \propto \int_{0}^{x} \frac{{\rm d} x}{v}.
\end{equation}
In the case of scale-free profile $P$ can be expressed as a function of a
single variable $u=x/x'$. The normalization of $P$ is provided by the
obvious condition $P(x, x'=x)=1$. The velocity $v$ is calculated by
integrating the equation of motion of the shell ${\rm d}^2 r/{\rm d}t^2 =
-G M/r^2$.

In the scale-free case it is possible to obtain analytical solutions for
$P$
\begin{eqnarray}
    P(u) &=& \frac{\gamma \Gamma[\gamma/(4-2\gamma)] u}{2 \sqrt{\pi}
    \Gamma[1/(2-\gamma)]}   \label{f4} \\
    & \times &
    \ _{2}F_{1} \left[\frac{1}{2},
    \frac{1}{2-\gamma},1+\frac{1}{2-\gamma}, u^{2-\gamma} \right]
    \nonumber
\end{eqnarray}
for $\gamma < 2$ and
\begin{eqnarray}
    P(u) &=& \frac{2 \Gamma[1/(\gamma-2)] u^{\gamma/2}}{\sqrt{\pi} \gamma
    \Gamma[\gamma/(2 \gamma-4)]}   \label{f5} \\
    & \times &
    \ _{2}F_{1}
    \left[\frac{1}{2}, \frac{\gamma}{2 \gamma - 4}, 1 + \frac{\gamma}{2
    \gamma-4}, u^{\gamma-2} \right]   \nonumber
\end{eqnarray}
for $\gamma > 2$. These results provide generalizations of the limiting
cases considered by Zaroubi \& Hoffman (1993). The collapse factor can
then be evaluated numerically for any $\gamma$ and $U$.

When the fiducial density profile is scale-dependent as in the case
of interest here the calculation of $P$ and $m_{\rm add}$ is more
complicated and has to be done numerically. The velocity in equation
(\ref{f3}) is
\begin{equation}    \label{f6}
    v \propto [E(x') - E(x)]^{1/2}
\end{equation}
where
\begin{equation}    \label{f7}
    E(x) \propto \int \frac{M(x)}{x^2} {\rm d} x = \int \frac{x_{\rm i}
    \Delta_{\rm i}^{2}}{1+\Delta_{\rm i}} \frac{{\rm d} x}{{\rm d} x_{\rm
    i}} {\rm d} x_{\rm i}
\end{equation}
where we used the expression for mass within $x_{\rm i}$
\begin{equation}    \label{f8}
    M(x) = M(x_{\rm i}) = \frac{4 \pi}{3} \rho_{{\rm crit},0} (x_{\rm i}
    R)^3 [1 + \Delta_{\rm i} (x_{\rm i})]
\end{equation}
and equation (\ref{p37}) with $f=1$. The cumulative density distribution
$\Delta_{\rm i}$ is taken in generalized form with cut-off, equation
(\ref{p35}).

Due to the complicated form of the integrand in the expression (\ref{f7})
for $E$ the integration cannot be performed analytically, but the
integrand can be exactly approximated by a polynomial in $x_{\rm i}$
making the integration possible. After changing variables from $x$ to
$x_{\rm i}$ in equation (\ref{f3}) the value of $P$ can be calculated
numerically for any $(x_{\rm i}, x_{\rm i}')$ pair. The added mass in
equation (\ref{f1}) can then be obtained by summing up the input from $j$
intervals between a given shell and the shell that is presently turning
around $x_{{\rm i}, ta}$ (the turn-around radius is obtained in a similar
way as the virial radius $x_{{\rm i}, v}$, from $t_{\rm u}=t_{\rm
coll}/2$, see the next Section)
\begin{eqnarray}
    m_{\rm add} (x_{\rm i}) &=& 4 \pi \rho_{\rm crit,0} R^3 (1+z_{\rm i})^3
    \sum_{j} \int_{x_{{\rm i},j}}^{x_{{\rm i},j+1}} y_{\rm i}^2 [1 +
    \delta_{\rm i}(y_{\rm i})] \nonumber \\
    & \times &
    P \left( x_{\rm i}, \frac{x_{{\rm i},j}+x_{{\rm
    i},j+1}}{2} \right) {\rm d} y_{\rm i}.  \label{f9}
\end{eqnarray}
The result for $f$ is sensitive to the number of steps taken in the
integration but usually $j=10$ is enough to obtain $f$ with two digit
accuracy for $x_{\rm i} > 1$.

After calculating the factor $f$ for a number of $x_{\rm i}$ values we
can approximate the function $f(x_{\rm i})$ by a polynomial in $x_{\rm i}$
and introduce it into equation (\ref{p37}). From equation (\ref{p36}) we
will then obtain the final density profile of a halo.

\section{Comparison with the universal profile}

NFW performed a series of $N$-body
simulations of structure formation in flat and open universe for different
scale-free power spectra of the form (\ref{p1}). They concluded that the
density profiles of dark matter halos are well fitted in all cases by a
formula
\begin{equation}    \label{p20}
    \frac{\rho(x)}{\rho_{\rm crit,0}} =
    \frac{\delta_{\rm char}}{(r/r_{\rm s})(1+r/r_{\rm s})^{2}}
\end{equation}
with a single fitting parameter $\delta_{\rm char}$, the characteristic
density. The so-called scale radius $r_{\rm s}$ is defined by
\begin{equation}     \label{p21}
    r_{\rm s} = \frac{r_{v}}{c}
\end{equation}
where $r_{v}$ is the virial radius i.e. the distance from the centre
of the halo within which the mean density is $v$ times the critical
density. $c$ in equation (\ref{p21}) is the so-called concentration
parameter, which is related to the characteristic density by
\begin{equation}    \label{p22}
    \delta_{\rm char} = \frac{v c^3}{3 [{\rm ln} (1+c) - c/(1+c)]}.
\end{equation}

NFW assume $v=200$ for all considered cosmological models, which is not
strictly correct. For $\Omega=1$ the exact prediction of the
spherical model for the ratio of the density to the critical density for
objects collapsing at the present epoch is $v=18 \pi^2 \approx 178$ (see
e.g. Lacey \& Cole 1993) so the natural choice would seem to adopt the
value $v=200$ in order to keep the assumptions as close as possible to
those of NFW. However, this value is inconsistent with the more advanced
SIM considered here: the value of $v$ can be calculated and it proves to
be much lower than 200: it is of the order of 30 for the small masses and
reaches 200 only for the largest halos. Fitting the NFW formula with
$v=200$ to the results of SIM produces a significant offset between the
fit and the SIM results, one therefore has to adopt the true value of $v$.
Then the problem of comparison with the simulations arises: what is the
relation between masses of halos determined from simulations with different
$v$? Fortunately, at the virial radii the density profiles of halos in the
simulations have slope $-3$ so the mass grows very slowly
(logarythmically) with $r$ and the masses determined with $v=200$ and
$v=30$ should be almost the same.

Since the measurements of the halo density profiles in the $N$-body
simulations of NFW are all done at the final epoch which is identified
with the present, in the following I will assume that for all the halos the
collapse time is the present epoch. According to the spherical
model the collapse time of the shell $x_{\rm i}$ in the $\Omega=1$
universe is (Padmanabhan 1993)
\begin{equation}    \label{p27}
    t_{\rm coll} = \frac{\pi (1+\Delta_{\rm i})}{H_{0}
    (1+z_{\rm i})^{3/2} \Delta_{\rm i}^{3/2}}.
\end{equation}
The present age of the universe is $t_{\rm u} = 2/(3 H_{0})$ and solving
equation $t_{\rm coll} = t_{\rm u}$ for $\Delta_{\rm i}$ gives us the values of
the cumulative density contrast as a function of $z_{\rm i}$, which will
be denoted by $\Delta_{{\rm i},v}$. This density contrast corresponds to
the presently collapsing initial shell $x_{{\rm i}, v}$ which can then be
found by numerically solving equation
\begin{equation}    \label{p28}
    \Delta_{\rm i,cut} (x_{\rm i}) = \Delta_{{\rm i},v},
\end{equation}
and the corresponding final virial radius $x_{v}$ is obtained from equation
(\ref{p37}). In order to solve this equation we have to specify the
initial conditions: the height of the peak $a$ and the rms fluctuation of
the density field $\sigma$. I have already chosen $a=3$ and will assume
that for all halos the starting point is such that $a \sigma=0.1$, small
enough for the linear theory to be valid. Final results are not very
sensitive to this particular choice.

We also need to make connection to the present magnitude of
fluctuations so I will adopt the standard normalization of the density
field such that $\sigma_{8} = 1$ (rms fluctuation in spheres of radius $8
h^{-1}$ Mpc is one). This normalization can be also expressed in terms of
the present nonlinear mass $M_{\ast}$ used by NFW. This mass is defined by
the condition of the rms fluctuation at this mass scale being equal to the
present critical threshold for collapse $\delta_{\rm crit}$. Given the
dependence of $\sigma$ on the smoothing scale this yields
\begin{equation}         \label{p38}
    M_{\ast} = M(R) \left[ \frac{\sigma_{TH}(R)}{\delta_{\rm crit}}
    \right]^{6/(n+3)}
\end{equation}
where $R$ is the smoothing scale used for normalization, subscript TH
refers to the top-hat smoothing, equation (\ref{a7}), and $M(R) = 4 \pi
R^3 \rho_{b}/3$. For $\Omega=1$ the threshold is $\delta_{\rm crit} =1.69$
so with our normalization at $R = 8 h^{-1}$ Mpc we have
\begin{equation}    \label{p39}
   M_{\ast} = 5.96 \ (1.69)^{-6/(n+3)} \times 10^{14} \ h^{-1} M_{\odot}.
\end{equation}

Once the normalization is set and the conditions $a=3$ and $a \sigma=0.1$
are adopted choosing the initial redshift $z_{\rm i}$ for a given spectral
index $n$ gives us the comoving smoothing scale $R$ with which the
overdense regions are identified. The mass of the halo within the virial
radius $x_{v}$ can then also be determined. According to the results of
the previous section the total mass inside the virial radius is
\begin{equation}    \label{p40}
    M = M(x_{{\rm i},v}) + m_{\rm add}(x_{{\rm i},v}).
\end{equation}

One of the main results of NFW was the dependence of the shape of
the density profiles of halos on their mass: their profiles were steeper
for lower masses. On the other hand, the standard prediction of SIM,
equation (\ref{p9d}), gives the same profile independently of mass.
However, with the improvements introduced in Section~3 and 4 it is
possible to reproduce the dependence of the profiles on mass.

It has been suggested in the literature (e.g. Katz, Quinn \& Gelb
1993) that $N$-body simulations indicate that if the density field is
smoothed with one scale $R$ lower peaks end up as galaxies and higher
ones as clusters. This, however, would violate the hierarchical way of
structure formation since higher peaks collapse earlier. Another argument
against such assumption comes from the calculations based on the improved
SIM for $n=-1$: the reasonable range of peak heights $a$ between 2 and 4,
which are most likely to produce halos, leads for a given smoothing scale
to the range of masses spanning only one order of magnitude, while in
$N$-body simulations NFW observe halos with masses spanning few orders of
magnitude. This suggests that the dependence on mass should rather be
related to the initial smoothing scale.

Calculations show that the mass of a halo increases with the
smoothing scale up to the scale for which the solution of equation
(\ref{p28}) yields rather small values of $x_{{\rm i}, v}$ (of the order
of unity). In these cases only the nearest neighbourhood of the peak
determined with the smoothing scale $R$ managed to collapse by the
present. Such peaks cannot be considered as collapsed structures because
they accreted up till now only a small fraction of the mass that is bound
to them (i.e. contained inside the cut-off scale). Therefore in the
following I will deal only with objects that have mass increasing with the
smoothing scale. For clarity hereafter I will consider different smoothing
scales for only one height of the peak $a=3$.

NFW suggest that the dependence of the shape of the halo on its mass is
related to the formation time of the halo: smaller halos that formed
earlier have steeper profiles. Although I have assumed the collapse time
to be the present epoch for all halos, this faster collapse of smaller
halos can nevertheless be observed: smaller masses are obtained
from smaller smoothing scales which correspond to higher initial redshifts
$z_{\rm i}$ and from equation (\ref{p27}) we have that constant
collapse time leads to lower $\Delta_{\rm i}$ and therefore higher
$x_{{\rm i}, v}$. Smaller halos form earlier in the sense that their
initial redshifts are higher and by the present epoch their virial radii
$x_{{\rm i}, v}$ reach the cut-off scale.

\begin{figure}
\begin{center}
    \leavevmode
    \epsfxsize=8cm
    \epsfbox[40 37 348 298]{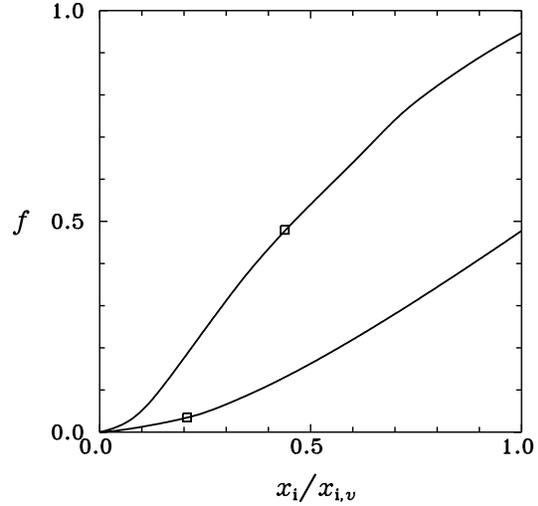}
\end{center}
    \caption{The collapse factor $f$ as a function of the initial radius
    of a shell (in units of the virial radius) for spectral index $n=-1$.
    The upper and lower curves correspond respectively to the small and
    large mass.}
\label{f}
\end{figure}

Given the initial conditions as specified above we may now proceed to the
direct comparison of the profiles obtained from SIM and the simulations.
An important step in determination of SIM profiles is the calculation of
the collapse factor (see Section~4). Figure~\ref{f} shows the dependence of
this factor on the initial radius of the shell for two different masses in
the case of $n=-1$. The upper curve in the figure corresponds to the
smallest mass available for $n=-1$ (with $z_{\rm i} = 1500$), i.e. $0.0017
M_{\ast} = 2.1 \times 10^{11} h^{-1} M_{\odot}$. The lower curve is for
the largest mass (with $z_{\rm i} = 43$), $3.3 M_{\ast} = 4.1 \times
10^{14} h^{-1} M_{\odot}$. Note that the initial radius was expressed in
units of the initial virial radius $x_{{\rm i},v}$, which is much larger
for the smaller mass. The squares marked in the plot show the points
corresponding to $0.01 x_{v}$. Since the density profiles will be
considered in the range $0.01 x_{v} < x < x_{v}$, only the collapse factor
values to the right of the marked points will be taken into account in the
calculations.

The behaviour of the collapse factor confirms the intuition one may obtain
from the scale-free case, equations (\ref{f2})-(\ref{f5}): the collapse
factor is smaller for higher effective index of the cumulative density
distribution (see the discussion in Zaroubi \& Hoffman 1993). Near the
center of the halo, where the initial density distribution is flat, the
outer shells contribute significantly to the inner mass and the contraction
is stronger. In more steeper parts of the density distribution the outer
shells contribute less and contraction is weaker. The dependence on mass
comes from the fact that small mass halos have large virial radii
$x_{{\rm i},v}$ reaching steeper parts of the distribution and even the
cut-off scale, while in large mass halos the final density distribution
comes from shells that initially were quite close to the peak.

\begin{figure*}
\begin{minipage}{18cm}
\begin{center}
    \leavevmode
    \epsfxsize=15cm
    \epsfbox[59 42 551 534]{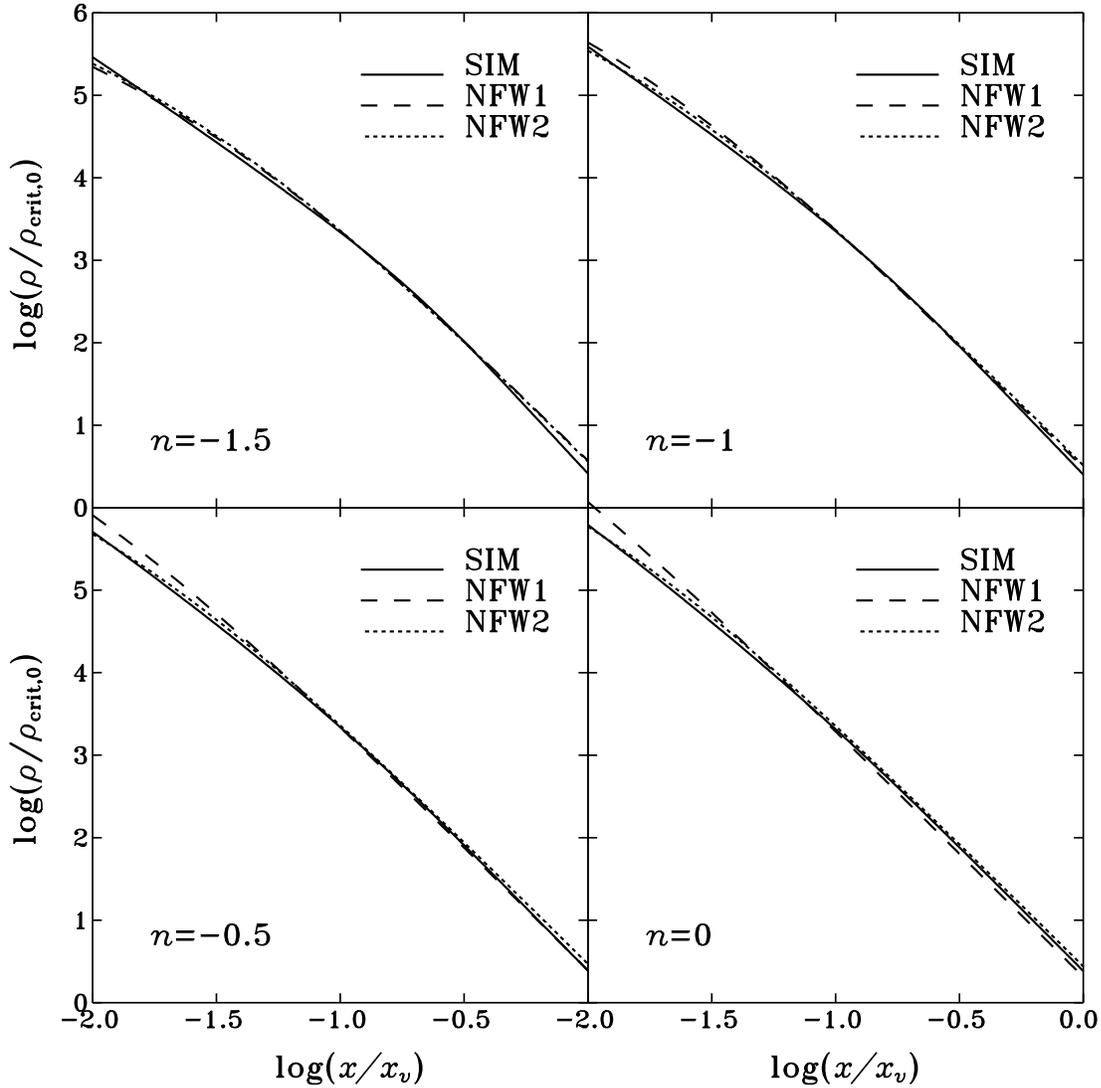}
\end{center}
    \caption{The density profiles of dark matter halos of mass of the
    order of $0.02
    M_{\ast}$ for different spectral indices $n$ in the range $0.01 x_{v} <
    x < x_{v}$. The solid lines show the predictions of SIM. The
    long-dashed ones give the NFW results with their fitted concentrations,
    while the short-dashed curves present the NFW formula (\ref{p25}) with
    concentrations fitted to SIM results.}
\label{s}
\end{minipage}
\end{figure*}

\begin{table*}
\begin{minipage}{18cm}
\caption{The parameters characterizing the small mass halos whose density
profiles are presented in Figure~\ref{s}.}
\label{small}
\begin{tabular}{rrcccrccrlc}
  $n$ & $z_{\rm i}$ & $\Delta_{{\rm i},v}$ & $R[h^{-1}$ Mpc] & $x_{{\rm
  i}, v}$ & $x_{v}$ & $r_{v}[h^{-1}$ Mpc] & $M[10^{12} h^{-1} M_{\odot}]$ &
  $M/M_{\ast}$ & $c_{\rm NFW1}$ &  $c_{\rm NFW2}$\\
  \hline
-1.5 & 350 & 0.00805 & 0.142 & 7.54 &  879 & 0.356 & 1.55 & 0.0211 & $\ \,
31.0$ & $ 33.5$ \\
-1.0 & 600 & 0.00469 & 0.188 & 6.76 & 1348 & 0.422 & 2.58 & 0.0208 & $\ \,
59.2$ & 46.9 \\
-0.5 & 1000& 0.00281 & 0.226 & 6.21 & 2065 & 0.466 & 3.45 & 0.0203 & 131 &
65.7 \\
 0.0 & 1500& 0.00187 & 0.268 & 5.73 & 2861 & 0.511 & 4.53 & 0.0212 & 306 &
84.4
\end{tabular}
\end{minipage}
\end{table*}

\begin{figure*}
\begin{minipage}{18cm}
\begin{center}
    \leavevmode
    \epsfxsize=15cm
    \epsfbox[59 42 551 534]{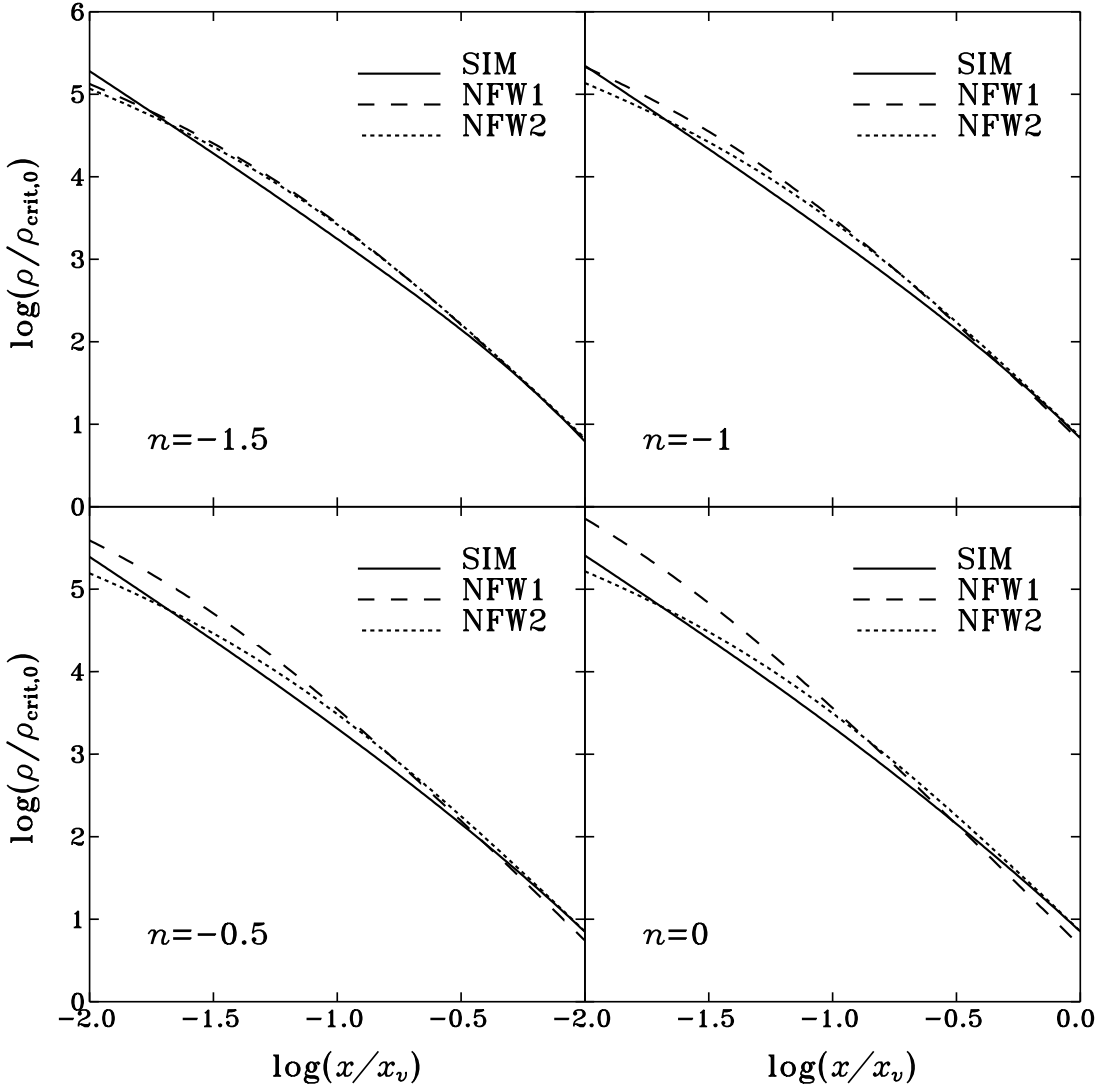}
\end{center}
    \caption{The same as Figure~\ref{s} but for masses of the order of
    $0.5 M_{\ast}$.}
\label{b}
\end{minipage}
\end{figure*}

\begin{table*}
\begin{minipage}{18cm}
\caption{The parameters characterizing the large mass halos whose density
profiles are presented in Figure~\ref{b}. }
\label{big}
\begin{tabular}{rccccrccrcc}
  $n$ & $z_{\rm i}$ & $\Delta_{{\rm i},v}$ & $R[h^{-1}$ Mpc] & $x_{{\rm
  i}, v}$ & $x_{v}$ & $r_{v}[h^{-1}$ Mpc] & $M[10^{13} h^{-1} M_{\odot}]$ &
  $M/M_{\ast}$ & $c_{\rm NFW1}$ &  $c_{\rm NFW2}$\\
  \hline
-1.5 & 150 & 0.0188 & \  0.438 & 6.50 & 301 & \  0.872 & 3.18 & 0.432 &
 15.6 & 14.0 \\
-1.0 & 180 & 0.0157 & \  0.625 & 5.45 & 298 & 1.03  & 5.55 & 0.447 & 21.6
& 15.1 \\
-0.5 & 200 & 0.0141 & \  0.816 & 4.67 & 280 & 1.14  & 7.83 & 0.461 & 34.9
& 16.2 \\
 0.0 & 200 & 0.0141 &    1.02  & 4.01 & 240 & 1.22  & 9.83 & 0.470 & 62.0
& 16.8
\end{tabular}
\end{minipage}
\end{table*}

\begin{figure*}
\begin{minipage}{18cm}
\begin{center}
    \leavevmode
    \epsfxsize=15cm
    \epsfbox[40 35 566 300]{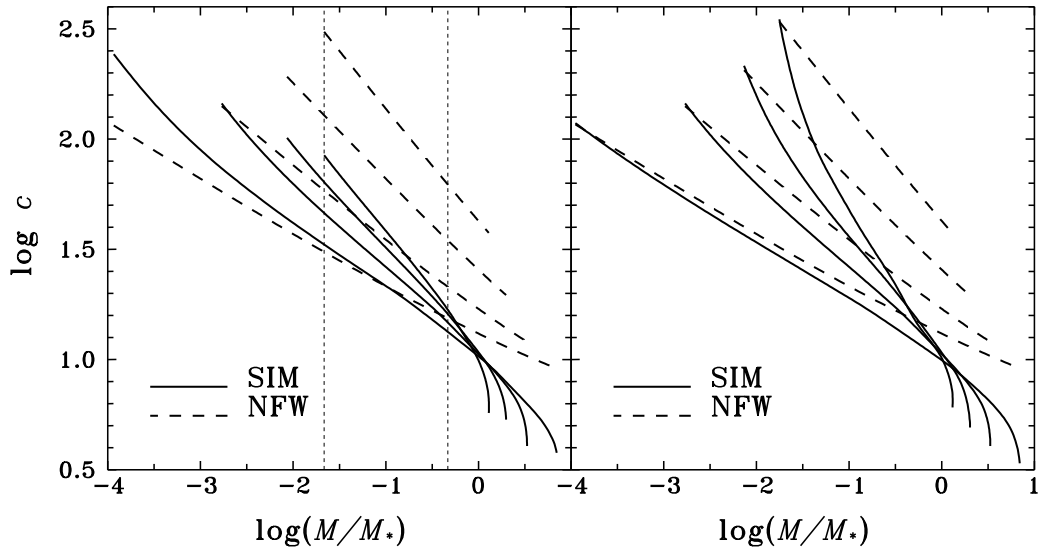}
\end{center}
    \caption{The dependence of the concentration parameter $c$ on
    the mass. The dashed lines show the results of $N$-body simulations of
    NFW, and the solid curves represent the results of SIM. In
    each set the curves from top to bottom correspond to $n=0,-0.5,-1$
    and $-1.5$ respectively. In the left panel the SIM results were
    obtained with constant $w=0.6$ while in the right panel the values
    were $w=0.14, 0.31, 0.6, 0.95$ respectively for $n=0,-0.5,-1$
    and $-1.5$.}
\label{kon}
\end{minipage}
\end{figure*}

In the comparison between the density profiles predicted by SIM to the
results of the $N$-body simulations of NFW I will use the NFW formula
(\ref{p20}) in the form
\begin{equation}    \label{p25}
    \frac{\rho(x)}{\rho_{\rm crit,0}} = \frac{v c^3}{3 [{\rm ln} (1+c) -
    c/(1+c)] (c x/x_{v}) (1+ c x/x_{v})^{2}}
\end{equation}
where definition of characteristic density (\ref{p22}) was used and the
distances were expressed in units of the smoothing radius and denoted by
$x$ in order to provide direct correspondence with the predictions of
SIM. NFW claim that their fitting formula is valid in the
range $0.01 x_{v} < x < x_{v}$. I therefore fit the formula (\ref{p25})
to the points ($x, \rho/\rho_{\rm crit,0}$) obtained from equations
(\ref{p36})-(\ref{p37}), spaced uniformly in log $(x/x_{v})$ in this
range. For a given spectral index $n$ the single fitting parameter is
the concentration $c$.

NFW consider density profiles of halos of mass roughly in the range $0.01
< M/M_{\ast} < 10$ so the initial redshifts will be chosen here so as to
obtain similar range of masses. Examples of profiles for halos of
different mass are shown in Figures~\ref{s} and \ref{b}. Figure~\ref{s}
shows the density profiles for objects of mass of the order of $0.02
M_{\ast}$, which corresponds to a big galaxy mass scale, while
Figure~\ref{b} shows the profiles for halos of mass of the order of
$0.5 M_{\ast}$, which is closer to the mass of a galaxy cluster. Such
choice of the mass values is motivated by the range of masses obtained for
different spectral indices (see Figure~\ref{kon} and the following
discussion). The initial redshifts needed to obtain halos of those masses
and all the following parameters are given in Tables~\ref{small} and
\ref{big} for the small and large masses respectively.

In agreement with the above discussion of the dependence of mass on the
smoothing scale, smaller masses require smaller smoothing scales and
the condition $a \sigma=0.1$ translates into their larger initial
redshifts $z_{\rm i}$. In the case of small masses the initial virial
radii $x_{{\rm i}, v}$ are of the order of the cut-off scale $x_{\rm
i,pp}/2$ which means that the cut-off really influences the formation of
small halos, i.e. in the absence of the cut-off their virial radii would
be much larger. In the case of the large mass halos the situation is
different: their virial radii $x_{{\rm i}, v}$ are significantly smaller
than the cut-off scale: this scale could not be reached by the present
epoch. The proper virial radii $r_{v} = x_{v} R/(1+z_{\rm i})$ are few
times larger for larger masses and in both cases agree well with the
observational estimates of the sizes of the halos of galaxies and
clusters.

The last two columns of Tables~\ref{small} and \ref{big} give the
concentrations $c_{\rm NFW1}$ and $c_{\rm NFW2}$ of the NFW formula. The
values of $c_{\rm NFW1}$ were calculated from the model based on
merging formalism provided by NFW that describes best the results of their
$N$-body simulations, while $c_{\rm NFW2}$ are the values of concentration
obtained by fitting formula (\ref{p25}) to the results of SIM. Following
the arguments presented above the fits were done for the value of $v$
calculated from the SIM. The width of the filter $w$ was chosen so that
the concentrations from NFW and SIM match for the smallest halos in the
$n=-1$ case.

Figure~\ref{kon} shows how the concentration $c$ depends on the mass of
the halo and the spectral index in the larger range of masses. The plots
cover the largest range of mass available from SIM for different spectral
indices with the initial conditions I adopted. The smallest masses were
obtained for $z_{\rm i}=1500$ and the largest correspond to the point
where the mass starts to decrease with $z_{\rm i}$ (the initial virial
radii are then rather small, significantly smaller than the cut-off
scale). The values of concentrations from the simulations were calculated
in the same range of mass as the one obtained from SIM.

The solid lines give the concentrations obtained from fitting SIM
results and the dashed lines are the $N$-body results of NFW (one has to
remember that those curves are fits to points displaying some scatter
that were themselves obtained by fitting the concentrations to the highly
irregular density profiles of halos in the simulations). In each set the
curves from top to bottom correspond to $n=0,-0.5,-1$ and $-1.5$
respectively. The results for higher spectral indices are shown in the
smaller range of mass because obtaining halos of smaller mass for example
in the case of $n=0$ would require initial redshift of $z_{\rm i} > 1500$
i.e. initial time earlier than the recombination epoch. Similar behaviour
-- more massive and more slowly forming halos for larger spectral indices
-- is observed in the $N$-body simulations of NFW. The two vertical dotted
lines in the left panel indicate the mass scales chosen for detailed
comparison shown in Figures~\ref{s}-\ref{b} and
Tables~\ref{small}-\ref{big}.

The left panel of Figure~\ref{kon} shows the SIM results for the constant
width of the cut-off filter $w=0.6$ which matches the concentrations for
smallest masses in the case of $n=-1$. There is however no reason why the
same width of filter should apply for different spectral indices so we can
try to adopt different widths for different spectral indices. Results of
matching the SIM and NFW concentrations for the smallest masses in all
cases are shown in the right panel of Figure~\ref{kon}. The adopted widths
were $w=0.14, 0.31, 0.6, 0.95$ respectively for $n=0,-0.5,-1$ and $-1.5$.
Although we have no way of estimating the exact shape of the cut-off
function from theory, one can expect that the cut-off will be sharper for
peaks with sharper initial density distribution i.e. for higher spectral
indices. This guess is in agreement with the dependence of the best
matching values of $w$ on the spectral index.

The predictive power of the improved SIM suffers also from the fact that
the discrepancy between the SIM and NFW concentrations is much more
sensitive to the unknown parameter $w$ than to the cut-off scale, which we
have estimated from the peaks formalism. Calculations of the profiles with
different cut-off scales show that e.g. decreasing the cut-off scale gives
smaller mass and steeper profile but because the mass is strongly affected
the corresponding concentration from the simulations is also decreased and
the factor by which both concentrations differ remains roughly the same.
On the other hand, decreasing the width of the cut-off filter does not
significantly affect the mass while the concentration of SIM profiles is
increased.

Leaving aside the exact determination of the width of the
filter, Figure~\ref{kon} proves that SIM predicts the same general trend in
the dependence of the shape of dark halo profiles on their mass: the
smaller the halo mass the steeper the profile. The dependence of the shape
of the halo on the initial power spectrum is also reproduced in the wide
range of masses: concentrations are larger for higher spectral indices up
to a mass of the order of the nonlinear mass $M_{\ast}$ where SIM
approximation seems to break down. We can also conclude from
Figure~\ref{kon} that the agreement between SIM and the simulations is
better for lower spectral indices. One may interpret this result as an
indication that for spectra with more large scale power and long-distance
correlations (low $n$) accretion is a realistic mechanism of growth of
fluctuations. On the other hand, in the case of higher spectral indices
isolated halos tend to appear and even relatively small halos form by
merging of yet smaller objects.

Another argument for the reliability of SIM comes from the possible (and
rather straighforward) application of the results presented here to more
realistic scale-dependent power spectra. Since smaller masses correspond
to smaller smoothing scales where the effective spectral index of a
realistic spectrum is lower, the dependence of the concentration on mass
according to the SIM should be flatter. This effect is also well visible in
the simulations of NFW: for the CDM spectrum the trend of smaller
concentrations with growing mass is preserved but the dependence on mass
is rather weak (especially when we take into account large scatter in
the results).

\section{Discussion}

The improved SIM presented in this paper provides simple understanding of
the dependence of the shape of the halo on its mass: smaller halos start
forming earlier and by the present epoch their virial radii reach the
cut-off scale that accounts for the presence of the neighbouring
fluctuations; more massive halos form later and their virial radii are not
affected by the cut-off scale, their virialized regions contain only the
material that initially was quite close to the peak identified with the
smoothing scale corresponding to the mass.

Although the concentrations of profiles predicted by SIM depend on the
exact shape of the cut-off function, this study leads to rather firm
conclusion that the agreement between the $N$-body and SIM profiles is
the better the smaller the mass of the halo and the lower the spectral
index of the initial power spectrum of fluctuations. If the profiles
obtained from the simulations indeed reflect reality this may indicate that
galactic halos form mostly by accretion, while for cluster size objects
merging must be taken into account. As suggested by Syer \& White (1998)
the universal profile can be formed by sufficiently dense satellites
reaching the centre of a halo intact and forming a cusp. Since the
attempts to obtain the dependence of the profiles on mass only from
formalisms describing the merging of halos (Nusser \& Sheth 1999,
Avila-Reese, Firmani \& Hernandez 1998) have not been fully successful, it
seems that the best description of halo profiles should be provided by a
model dealing with a combination of accretion and merging (for the
discussion on the distinction between the two phenomena see
Salvador-Sol\'{e}, Solanes \& Manrique 1998). This work provides the
understanding of the origin of universal profile in the case of small
masses for which SIM can be considered a valid approximation.

One of the important shortcomings of SIM, that was
not mentioned here, is the artificial combination of the linear and
strongly nonlinear regime without taking into account the weakly nonlinear
phase that may affect the initial density distribution before the strongly
nonlinear evolution takes over. As discussed by \L okas (1998) these
corrections are of similar importance as the ones introduced by
using the formalism of BBKS to describe peaks in the density field
instead of overdense regions. Both effects tend to steepen the initial
density profiles but are difficult to model analytically (for the
corrections for peaks see e.g. Ryden 1988). With the modifications of
SIM introduced here we are interested mostly in regions
not very distant from the centre of the forming halo. In these regions
weakly nonlinear corrections to the expected overdensity $\langle \delta
\rangle$ are known only numerically and it is difficult to obtain the
cumulative density $\langle \Delta_{\rm i} \rangle$. Even if we
approximate it by some analytical expression we cannot proceed because the
formula for the final profile is so complicated that any perturbative
treatment is impossible. Qualitatively one may expect that the final
profile will be somewhat steeper but since the value of $\Delta_{{\rm
i},v}$ will not be changed the solution for the virial radius $x_{{\rm i},
v}$ will be lower. It follows that also the halo mass will be decreased
but it will have to be compared to a less massive and therefore steeper
halo from the simulations, so it is difficult to predict whether the
agreement between the two results would be improved. It should be added
that this picture of weakly nonlinear corrections does not take into
account the parallel evolution of the rms fluctuation $\sigma$ itself. As
discussed by \L okas et al. (1996) its value in the weakly nonlinear
regime may be close to linear in the case of $n=-1$ but may differ
significantly from the linear prediction for other spectral indices.

No solution to the problem of dark halo formation cannot be considered
valid without a detailed agreement between its predictions and
observations. To date several such comparisons were performed, in most
cases in the form of fitting the NFW profile to the observed profiles of
galaxies and clusters that are believed to be dominated by dark matter or
provide some indication on how light traces mass. Carlberg et al.
(1997) find that universal profile of NFW provides a very good fit to the
density profiles of clusters in the CNOC survey, while Adami et al.
(1998) find that galaxy distribution in clusters in the ENACS
sample displays a core rather than a cusp in the central regions, but the
mass distribution (Adami, private communication) is somewhat steeper.
These results suggest that the universal profile indeed provides a good
description of the density distribution in clusters. In the case of
galaxies the situation is less satisfying. Kravtsov et al. (1998) analysed
density profiles of dwarf and low surface brightness galaxies and found
that they are much better fitted by a so-called Burkert profile (Burkert
1995) with a flatter cusp than in the NFW formula. They also performed a
series of $N$-body simulations based on a different code than that used by
NFW and found similar shapes of galaxies as those observed.

\section*{Acknowledgments}
This work was supported in part by the Polish State Committee for
Scientific Research grant No. 2P03D00815 and the French Ministry of
Research and Technology within the program Jumelage (Astronomie Pologne).
I wish to thank A. Mazure for his hospitality at Laboratoire
d'Astronomie Spatiale in Marseille where part of this work was done.

\appendix
\section{The scale of coherence}

The calculation of the scale of coherence $x_{\rm i,COH}$ of the cumulative
density contrast $\Delta_{\rm i}$ can be done in a way analogous to the
calculation of the scale of coherence $x_{\rm coh}$ of the density contrast
$\delta$ itself (\L okas 1998). The condition for $x_{\rm coh}$ was
\begin{equation}   \label{a1}
    \langle \delta \rangle = \langle (\delta - \langle \delta \rangle)^2
    \rangle^{1/2}
\end{equation}
where
\begin{equation}      \label{a2}
    \langle \delta \rangle = a \sigma \varrho
\end{equation}
\begin{equation}        \label{a3}
    \langle (\delta - \langle \delta \rangle)^2 \rangle = \sigma^2
    (1- \varrho^2).
\end{equation}
The results were obtained from the two-point Gaussian probability
distribution function for fields $\delta$ and $\gamma$, from which a
conditional probability distribution for $\delta$ followed given that at
distance $r$ from where $\delta$ is measured there is an overdense region
where the density is $\gamma = a \sigma$.

Here we may proceed in a similar fashion and consider a two-point
probability distribution for variables $\Delta_{\rm i}$ given by equation
(\ref{p6}) and $\gamma$. If the condition for $\gamma$ is the same we have
(see also HS, Peebles 1984)
\begin{equation}          \label{a4}
    \langle \Delta_{\rm i} \rangle = a \Sigma \varrho'
\end{equation}
\begin{equation}            \label{a5}
    \langle (\Delta_{\rm i} - \langle \Delta_{\rm i} \rangle)^2 \rangle =
    \Sigma_{\Delta}^2 = \Sigma^2 (1- \varrho'^2).
\end{equation}
The symbols $\Sigma$ and $\Sigma_{\Delta}$ refer respectively to the
unconstrained and constrained dispersions of the $\Delta_{\rm i}$ field.
The new correlation coefficient $\varrho'=\xi'/(\sigma \Sigma)$ is the
normalized correlation function $\xi'(r_{\rm i})=\langle \Delta_{\rm
i}(r_{\rm i}) \gamma(0) \rangle$.

The definition of $\Delta_{\rm i}$, equation (\ref{p6}), can be rewritten
as
\begin{equation}          \label{a6}
    \Delta_{\rm i} (r_{\rm i}) = \frac{1}{(2 \pi)^3} \int \delta(k)
    W_{\rm TH}(k r_{\rm i}) {\rm d}^3 k
\end{equation} where $W_{\rm TH}(k r_{\rm i})$ is the top-hat window
function
\begin{equation}          \label{a7}
    W_{\rm TH}(k r_{\rm i}) = \frac{3 [\sin (k r_{\rm i}) - k r_{\rm i}
\cos (k r_{\rm i})]}{(k r_{\rm i})^3}.
\end{equation}
Expression (\ref{a6}) leads to
\begin{equation}          \label{a8}
    \langle \Delta_{\rm i} (r_{\rm i}) \rangle = \frac{a}{(2 \pi)^3 \sigma}
    \int P_{R}(k) W_{\rm TH}(k r_{\rm i}) {\rm d}^3 k
\end{equation}
where $P_{R}(k)$ is the power spectrum initially smoothed with the
Gaussian filter (\ref{p2})
\begin{equation}          \label{a9}
    P_{R}(k) = P(k) {\rm e}^{-k^2 R^2}.
\end{equation}
The expected cumulative density is therefore
\begin{equation}          \label{a10}
    \langle \Delta_{\rm i} \rangle = \frac{6 C a}{(2 \pi)^2 r_{\rm i}^{n+3}
    \sigma} I_{1}(x_{\rm i})
\end{equation}
where
{\samepage
\begin{eqnarray}
    I_{1}(x_{\rm i}) &=& \int_{0}^{\infty} k^{n-1} (\sin k - k \cos k) \
    {\rm e}^{-k^2/x_{\rm i}^2} {\rm d} k  \nonumber \\
    &=& \frac{1}{2} x_{\rm i}^{n+1} \Gamma \left( \frac{n+1}{2} \right)
    \left[ _{1} F_{1} \left( \frac{n+1}{2}, \frac{3}{2}, -
    \frac{x_{\rm i}^2}{4} \right) \label{a11} \right.\\
    &-& \left. _{1} F_{1} \left( \frac{n+1}{2}, \frac{1}{2},
    -\frac{x_{\rm i}^2}{4} \right) \right]  \nonumber
\end{eqnarray}
}
Together with the expression for $\sigma$, equation (\ref{p4}), this leads
to the expression (\ref{p7}) for the expected cumulative density.

The unconstrained variance of $\Delta_{\rm i}$ is
\begin{equation}          \label{a12}
    \Sigma^2 = \frac{1}{(2 \pi)^3} \int P_{R}(k) W_{\rm TH}^{2}
    (k r_{\rm i}) {\rm d}^3 k = \frac{18 C}{(2 \pi)^2 r_{\rm i}^{n+3}}
    I_{2} (x_{\rm i})
\end{equation}
where
{\samepage
\begin{eqnarray}
    I_{2}(x_{\rm i}) &=& \int_{0}^{\infty} k^{n-4} (\sin k - k \cos k)^2 \
    {\rm e}^{-k^2/x_{\rm i}^2} {\rm d} k \label{a13} \\
    &=& \frac{1}{2} x_{\rm i}^{n-3} \Gamma \left( \frac{n-1}{2} \right)
    \left[ \frac{x_{\rm i}^2 + 2}{2} \ _{1} F_{1} \left( \frac{n-1}{2},
    \frac{1}{2}, -x_{\rm i}^2 \right) \right.  \nonumber \\
    &-& \left. \frac{n-2}{n-3} \ _{1} F_{1} \left( \frac{n-3}{2},
    \frac{1}{2}, -x_{\rm i}^2 \right) + \frac{x_{\rm i}^2}{2} +
    \frac{1}{n-3} \right]  \nonumber
\end{eqnarray}
and the correlation coefficient is given by
\begin{equation}          \label{a14}
    \varrho'^{2} = \frac{2}{\Gamma[(n+3)/2]} x_{\rm i}^{-(n+3)}
    \frac{I_{1}^2}{I_{2}}.
\end{equation}
}

Given the analytical expressions for $\langle \Delta_{\rm i} \rangle$,
$\Sigma$ and $\varrho'$ equation (\ref{p31}) has to be solved numerically
to determine $x_{\rm i} = x_{\rm i,COH}$. The results for different power
spectra are shown in Figure~\ref{ccohcpp}.

\end{document}